\documentclass[aps,pre,reprint,groupedaddress,amsmath,amssymb]{revtex4-1}

\usepackage{graphicx}
\usepackage{dcolumn}
\usepackage{bm}
\usepackage{enumitem}
\usepackage{amsmath}
\usepackage{amssymb}
\usepackage{subfigure}
\usepackage{xcolor}
\usepackage{float}

\begin{document}
\newcommand{\TriDown}{$T^{-}$}
\newcommand{\TriUp}{$T^{+}$}
\newcommand{\AsUp}{$A^{+}$}
\newcommand{\AsDown}{$A^{-}$}
\newcommand{\eqn}[1]{(\ref{#1})}
\newcommand{\ang}[1]{{\langle\, #1 \,\rangle}}
\newcommand{\mat}[1]{{\bf #1}}


\title{Analyzing Network Reliability Using Structural Motifs}


\author{}
\affiliation{}


\author{Yasamin Khorramzadeh}
\email{yasi@vbi.vt.edu}
\affiliation{Network Dynamics and Simulation Science Laboratory, Virginia Bioinformatics Institute, Virginia Tech, Blacksburg, Virginia 24061, USA}
\affiliation{Department of Physics, Virginia Tech, Blacksburg, Virginia 24061, USA}

\author{Mina Youssef}
\email{myoussef@vbi.vt.edu}
\affiliation{Network Dynamics and Simulation Science Laboratory, Virginia Bioinformatics Institute, Virginia Tech, Blacksburg, Virginia 24061, USA}

\author{Stephen Eubank}
\email{seubank@vbi.vt.edu}
\affiliation{Network Dynamics and Simulation Science Laboratory, Virginia Bioinformatics Institute, Virginia Tech, Blacksburg, Virginia 24061, USA}
\affiliation{Department of Physics, Virginia Tech, Blacksburg, Virginia 24061, USA}
\affiliation{Department of Population Health Sciences, Virginia Tech, Blacksburg, Virginia 24061, USA}

\author{Shahir Mowlaei}
\email{shahir@vbi.vt.edu}
\affiliation{Department of Physics, Virginia Tech, Blacksburg, Virginia 24061, USA}
\affiliation{Network Dynamics and Simulation Science Laboratory, Virginia Bioinformatics Institute, Virginia Tech, Blacksburg, Virginia 24061, USA}


\begin{abstract}
This paper uses the reliability polynomial, introduced by Moore and Shannon in 1956, to analyze the effect of network structure on diffusive dynamics such as the spread of infectious disease. 
We exhibit a representation for the reliability polynomial in terms of what we call {\em structural motifs} that is well suited for reasoning about the effect of a network's structural properties on diffusion across the network.
We illustrate by deriving several general results relating graph structure to dynamical phenomena.
\end{abstract}


\maketitle

\section{Introduction}

Characterizing networks in a way that is directly relevant to diffusion phenomena on the network is important, but difficult.
We argue that the \emph{Network Reliability Polynomial} introduced by Shannon and Moore~\cite{Moore:56}
 is a characterization that folds together static measures like degree, modularity, and measures of centrality into precisely the combinations that are most relevant to the dynamics.~\cite{DynamicNewman}
Conversely, knowledge of reliability can be used to infer structure, in the sense of network tomography.~\cite{Dynamic}
Furthermore, reliability is a useful concept for reasoning more generally about the consequences of structural changes. 
Colbourn~\cite{Colbourn} gives a comprehensive introduction to network reliability, while Youssef~\cite{Youssef:13} provides a brief derivation of the form of the reliability polynomial.

A network's {\em reliability} is the probability that it continues to function after sustaining damage to its component edges and/or vertices.
Reliability depends on a parameterized {\em damage model} $D(\vec{x})$, which specifies the probability of damaging a particular set of components,
and a parameterized property ${\cal P}(\vec{\alpha})$, which specifies what it means for the network to function.
Canonical problems in diffusion over fixed networks can be cast in terms of reliability.
For example, a damage model $D(x)$ under which edges (respectively, vertices) fail independently with probability $1-x$ represents bond (respectively, site) percolation.
The choice of ${\cal P}$ then selects the dynamical phenomenon of interest.
For example, the property ``contains a connected component including at least a fraction $\alpha$ of the vertices'' is appropriate for studying the existence of a giant component. 
For ease of notation, we often express ${\cal P}$ in terms of a corresponding {\em reliability rule} $r_{\cal P}$, a binary function indicating whether property ${\cal P}$ holds for a given graph.
We say that the rule {\em accepts} a graph $g$ if and only if $r(g) = 1$, that is $g$ has the property ${\cal P}$.
\color{black}

Here we introduce four different reliability rules. The first three are the most commonly used rules, followed by the last one that could be of use to study percolation problems. To make it easier to understand, we refer to real world problems like designing reliable communication networks and epidemiology which benefit from each of these rules:
\color{black}
\begin{enumerate}
\item {\em Two-Terminal:} a graph is accepted if it contains at least one directed path from a distinguished vertex $S$ (the {\em source}) to another distinguished vertex $T$ (the {\em terminus}). Reliability under this rule is the probability that the specified source can send a message to the specified terminus in a damaged network.~\cite{TwoTermCite}
\item {\em $K$-Terminal:} a graph is accepted if every vertex is in a connected component that includes at least one of $K$ specified vertices.
For example, consider a set of $K$ nodes as service centers. Then reliability under this rule is the probability that every operational site in a damaged network is connected to at least one service center.~\cite{KTermCite}
\item {\em All-Terminal:} a graph is accepted if it consists of a single connected component.
The reliability under this rule, also known as  {\em system reliability} is  the probability that every pair of nodes in a damaged network can communicate with each other, or alternatively, that any vertex in a damaged network can broadcast to all the other vertices.~\cite{AllTermCite}
\item {\em $EAR$-$\alpha$:}
To understand this rule we discuss the application of bond percolation for the study of the spread of infectious diseases on networks. Such a bond percolation disease model was explained by ~\cite{ PercolationEpidemiology} .
\color{black}
The probability that an edge does not fail represents the {\em transmissibility}, i.e., the conditional probability of transmitting infection from one person to another, conditioned on the source being infectious and the destination being susceptible.
One of the most important properties of disease dynamics is the {\em attack rate}, defined as the fraction of the population infected in an outbreak.
Most models of infectious disease exhibit a sharp transition in the attack rate at a critical value of transmissibility.
Indeed, this is a percolation phase transition.
Using reliability we can find the relationship between critical transmissibility and all structural information of a network contained in its edge list. 
\color{black}
In this case the reliability is the probability that the expected attack rate for an outbreak seeded in a single person chosen uniformly at random from the population is at least $\alpha$.
Hence, we call this rule $EAR$-$\alpha$.
\color{black}
\end{enumerate}
These rules are all {\em coherent}.
That is, any graph formed by adding an edge to an accepted graph is also accepted.
\color{black}
In a companion to this paper \cite{Youssef:13}, we have shown how the concept of network reliability together with an efficient, scalable estimation scheme can shed light on complicated dynamical trade-offs between local structural properties such as assortativity-by-degree and the number of triangles.
Here we introduce a different representation of the reliability polynomial that highlights the role certain network structures play in dynamical phenomena.
\color{black}
We show how coefficients of the reliability polynomial can be interpreted in terms of 
topological motifs in the network
\color{black}
and their overlaps.
Conversely, we illustrate how knowledge of these motifs and their overlaps can be used to infer important constraints on the dynamics of diffusion processes on the network.
The representation in \cite{Youssef:13} is well-suited for computational analysis of networks with up to $10^8$ edges, but is analytically tractable only for small networks;
the representation presented in the current work is analytically tractable, but computationally feasible only for small networks because of its combinatorial complexity.
Thus the results of this paper exactly complement those of the previous paper.
\color{black}

\section{Reliability polynomials}

We use the common notation of $G(V,E)$ for a graph with $V$ vertices and $E$ edges. 
The graph may be directed or undirected, and it is possible to have multiple edges between two vertices. 

The vertices and edges may be labeled.
The general case of directed edges and labeled vertices and edges is powerful enough to represent extremely complex networks such as interdependent infrastructure networks.
Here, without loss of generality, we restrict ourselves to homogeneous networks represented as undirected, unlabeled graphs.

\subsection{Definition and a common representation}

The {\em reliability} $R(G, {\cal P}(\vec{\alpha}), D(\vec{x}))$ of a network $G$ with respect to the property ${\cal P}$ under damage model $D$ is the probability that a subgraph of $G$ chosen with probability given by $D$ has property ${\cal P}$, the binary rule $r_{\cal P}$ examines whether subgraph has property $\cal P$ or not, if it has then: 
  $$ r_{{\cal P}(\vec{\alpha})}(g)=1, $$ otherwise $$ r_{{\cal P}(\vec{\alpha})}(g)=0.$$
 i.e. , reliability can be interpreted as the expected value of the reliability rule operator over different subgraphs of $G$. 
We will explicitly include the dependence on the network $G$ and the property ${\cal P}$ in notation such as $R(G, {\cal P}(\vec{\alpha}), D(\vec{x}))$  only when we wish to distinguish the reliability of two different graphs or two different properties.
Moreover, we will not include the damage model itself, but only the values of its parameters $\vec{x}$.
Finally, for a homogeneous network in which all edges (or all vertices) fail with the same probability, $\vec{x}$ is a scalar, $x$.
Thus we can write the reliability simply as $R(x)$.
\color{black}

\begin{eqnarray}
\label{eq:xpoly1}
R(x) \equiv \sum_{g \subseteq G} r_{{\cal P}(\vec{\alpha})}(g) p_{D(\vec{x})}(g) 
\end{eqnarray}
For the independent-edge damage model, in which the probability of selecting a subgraph $g \subseteq G$ depends only on the number of its edges, $|g|=k$, and is $x^k(1-x)^{E-k}$, we have:
\begin{eqnarray}
\label{eq:xpoly2}
R(x) &=& \sum_{g \subseteq G} r_{{\cal P}(\vec{\alpha})}(g) x^k(1-x)^{E-k}
\end{eqnarray}
We can re-write ~\ref{eq:xpoly2}in terms of sum over subgraphs of different sizes, introduced by Alon et al.\cite{ AlonNatRevGen} as \textit{motifs}:
\begin{eqnarray}
\label{eq:xpoly3}
 R(x)= \sum_{k=0}^E R_k  x^k(1-x)^{E-k}.
\end{eqnarray}
    
$R_k$ is the number of subgraphs of $G$ with exactly $k$ edges that are accepted by the rule.
For computational convenience, we often prefer to work with normalized coefficients 
\begin{equation}
\label{eq:Pdef}
P_k \equiv R_k / {E \choose k}.
\end{equation}
$P_k$ is the {\em fraction} of subgraphs of k$G$ with exactly $k$ edges that are accepted by the rule.
$P_k \le P_{k+1}$ for a coherent rule.
$P_k$ can be estimated efficiently via monte Carlo simulation. \cite{Youssef:13}

 Substituting $R_k$ coefficients in Equation ~\ref{eq:xpoly2} with $P_k$s from ~\ref{eq:Pdef}, we can see the resemblance to binomial distribution, since $P_k \le 1$ it is clear that 
\begin{eqnarray}
\label{eq:Rdomain}
R(x)  &=& \sum_{k=0}^E \binom{E}{k} P_k  x^k(1-x)^{E-k}  \nonumber \\
         &\le& \sum_{k=0}^E  \binom{E}{k} x^k(1-x)^{E-k} \le 1 
\end{eqnarray}
\color{black} Therefore, $R(x)\colon [0,1] \to [0,1]$ is a continuous polynomial with only a finite, but possibly large, number of coefficients $R_k, k \in \{0,\ldots, E\}$.
That is, the reliability can be thought of as a vector in an $E+1$-dimensional vector space, and the $R_k$'s as the components of the vector in the basis $x^k (1-x)^{E-k}$.
There are, of course, many other bases we could choose for this space.
An orthogonal basis, such as the first $E+1$ Legendre polynomials, might have useful estimation properties.
Here we use another non-orthogonal basis -- the functions $x^k$ \i.e. the Taylor series expansion -- because of its simplicity and its attractive interpretation. 
There is a unique mapping from coefficients in one basis to those in the other, which can be derived by expanding the factor $(1-x)^{E-k}$ in Equation~\ref{eq:xpoly3}:
\begin{eqnarray}
\label{eq:Npoly}
R(x) &=& \sum_{k=0}^{E} R_k x^k \sum_{m=0}^{E-k} {E-k \choose m} (-1)^m x^m \nonumber \\
        &=& \sum_{k=0}^{E} R_k \sum_{l=k}^{E}  {E-k \choose l-k} (-1)^{l-k} x^{l} \nonumber \\
        &=& \sum_{l=0}^{E} (-1)^l x^l \sum_{k=0}^{l}  (-1)^{k} R_k {E-k \choose l-k}  \nonumber \\
        &=& \sum_{l=0}^E N_l x^l
\end{eqnarray}
where
\begin{equation}
\label{eq:Ndef}
N_l \equiv (-1)^l \sum_{k=0}^{l} (-1)^{k} {E-k \choose l-k} R_k .
\end{equation}
The $N_l$ coefficients are signed integers.  In section ~\ref{sec:in-ex} we will explain how we can interpret these coefficients.
\color{black}

\subsection{Structural motifs}
\label{sec:st_motifs}
We can express the reliability polynomial in terms of overlaps among certain distinguished subgraphs.
These subgraphs are the ${\cal P}$-minimal subgraphs of $G$.
A graph $g$ is ${\cal P}$-minimal if and only if:
\begin{enumerate}
\item $g$ has property ${\cal P}$; and
\item there is no proper subgraph $g'\subset g$ that has property ${\cal P}$.
\end{enumerate}
Obviously, whether a graph is ${\cal P}$-minimal or not depends on the property ${\cal P}$, or equivalently here, the reliability rule.
For example, ${\cal P}$-minimal graphs under a Two-Terminal rule are paths from $S$ to $T$ with no extraneous edges, i.e., no loops or dead ends;
under the All-Terminal rule, they are spanning trees.
In general, the rule selects a distinctive topological pattern (e.g.\ path, spanning tree) that may occur many times in a given graph, i.e., if we consider all subgraphs of a distinctive pattern to be the {\em motif}s introduced by Alon et al.\cite{ AlonNatRevGen} then reliability rule selects a subset of these motifs that has the property $\cal P$.
We refer to subsets generated by a particular rule as {\em structural} motifs because, as we will demonstrate, they are the structural elements of the network that completely determine the occurrence of dynamical phenomena of interest, as specified by ${\cal P}$. One advantage of using this representation is that the contribution of the structural motifs to the reliability is known exactly for all ${\cal P}$ and all $D$, as is shown below.\par

\begin{figure}
\includegraphics[width=3in]{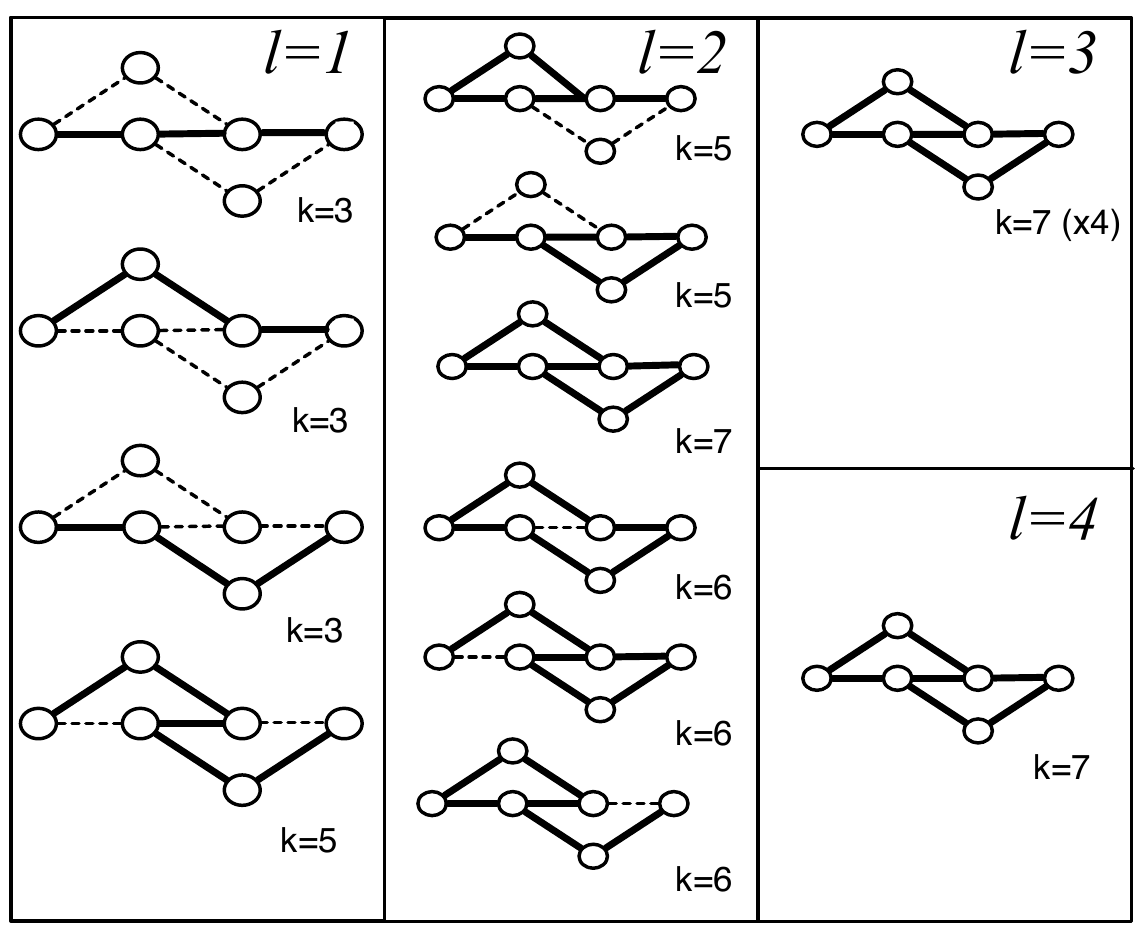}
\caption {Example calculation of two-terminal reliability in the toy network in the left panel. The second panel from the left shows the four motifs; the third shows all unions of two motifs. There are ${4 \choose 3} = 4$ unions of 3 motifs, all of which give the entire graph as shown on the top of the right panel; there is one union of all motifs, which is also the entire graph. The number next to each union of motifs gives its size. }
\label{fig:subgraph3}
\end{figure}

\color{black}
To describe this, we apply two terminal reliability rule with \textit{source} and \textit{terminus} nodes as its parameters on a small toy network, depicted in Figure~\ref{fig:subgraph3}. For this rule structural motifs are simple paths (including but not limited to shortest paths) connecting these two nodes. The structural motifs and their overlap can be seen in Figure~\ref{fig:subgraph3}.\par

\begin{figure}
\includegraphics[width=3in]{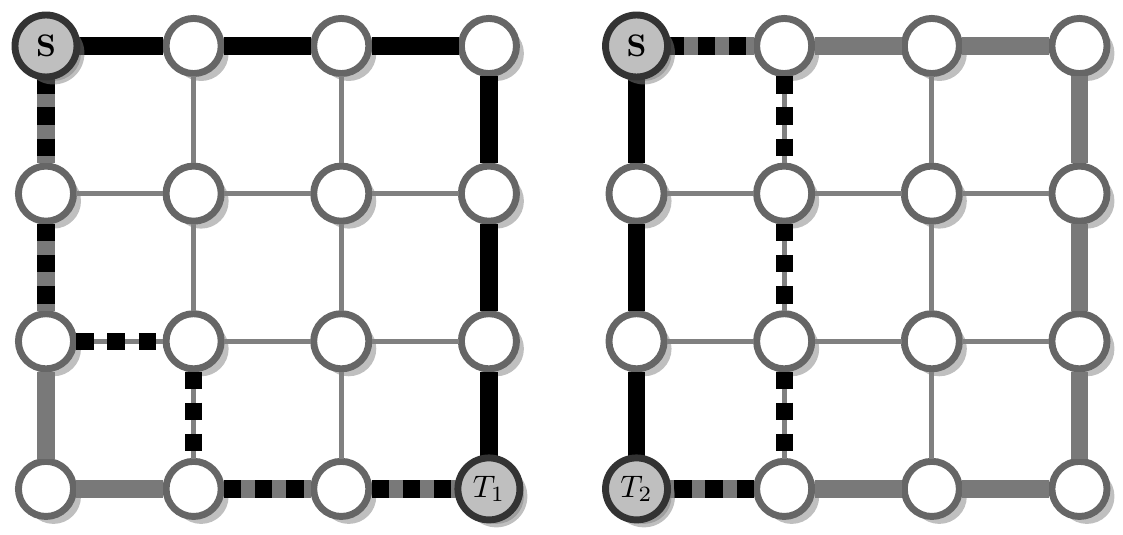}
\caption {Structural motif for the two-terminal reliability rule are shown for same source node $S$ and two different target nodes $T_1$ and $T_2$. We can see that structural motifs of the same network varies with the parameters of property $\cal P$. In the left grid three structural motifs of sizes 6 are shown: note that black dashed path and gray line have all but two edges in common. For the grid on the right, black normal, black dashed and gray paths represent motifs of sizes 3, 5, 9 respectively. The last two have two edges in common. }
\label{fig:2dgrids}
\end{figure}

In another example, we consider the two-terminal reliability rule on a two dimensional grid with 4 nodes in each dimension. For a given source node we look at two different terminus nodes to illustrate the dependence of structural motifs on parameters of the reliability rule.  Several structural motifs of different sizes for each set of selected sources and termini are shown in Figure~\ref{fig:2dgrids}.

\subsection{Contribution of structural motifs to $R(x)$}
\label{sec:in-ex}
The sizes of each motif and their unions completely determine a network's reliability. 
We demonstrate this using an Inclusion-Exclusion argument motivated by a series of straightforward examples.
A detailed proof and more examples are provided in Appendices A and B.\color{black}
\subsubsection{\label{sec:1motif}Example 1: A single structural motif}
Suppose the network $G$ contains only one structural motif and that it is a set of $k_0$ edges.
For example, for Two-Terminal reliability, suppose that there is exactly one path between $S$ and $T$, and that it has length $k_0$.
Then the motif will occur exactly once among all subgraphs of size $k_0$.
For $k>k_0$, we must ``use up'' $k_0$ edges to build the structural motif.
This leaves $E-k_0$ other edges from which, because the rule is assumed to be coherent, any set of $k-k_0$ produces an acceptable subgraph of size $k$.
Hence for this case
\begin{equation}
R_k = \left\{
\begin{array}{lr}
0 & k < k_0 \\
{E-k_0 \choose k-k_0} & k \ge k_0
\end{array}
\right.
\end{equation}

\subsubsection{\label{sec:Nmotifs}Example 4: $N$ disjoint structural motifs}
Suppose the graph has exactly $N$ structural motifs, that all have $k_0$ edges, and that the $N$ edge sets are disjoint.
Arguing as above, with the convention that ${a \choose b} = 0$ $\forall b<0$,  gives: 
\begin{equation}
R_k = \sum_{i=1}^{N} (-1)^{i+1} {N \choose i} {E - ik_0 \choose k - ik_0} 
\end{equation}

\subsubsection{\label{sec:2omotifs}Example 5: Two overlapping structural motifs}
Suppose the graph has exactly two structural motifs, that both have $k_0$ edges, and that the number of edges in the union of the two is  $k_0 +\Delta$.
Arguing as in Example~1, we get a similar result, with $2k_0$ replaced by $k_0+\Delta$:
\begin{equation}
R_k = \left\{
\begin{array}{lr}
0 & k < k_0 \\
2{E-k_0 \choose k-k_0} & k_0 \le k < k_0 + \Delta \\
2{E-k_0 \choose k-k_0} - {E-k_0 -\Delta \choose k-k_0-\Delta} & k_0 + \Delta \le k
\end{array}
\right.
\end{equation}

\subsubsection{\label{sec:genMotifs}The general case}
Suppose the graph has exactly $N$ structural motifs.
As above, its reliability polynomial will be determined by the size of each structural motif and the overlaps among them. 
Define $N^{(l)}_k$ as the number of combinations of $l$ structural motifs whose union contains exactly $k$ edges. 
Also, define 
\begin{equation}
N_k \equiv \sum_{l=1}^N (-1)^{l+1} N^{(l)}_k.
\label{eq:Nk}
\end{equation}
Then arguing as above gives
\begin{equation}
R_k = \sum_{k'=0}^{k} N_{k'} {E-k' \choose k-k'}.
\label{eq:Nk2Rk}
\end{equation}

In Appendix C, we present constraints on $N_{k}$.
Using Equation~\ref{eq:Nk} and ~\ref{eq:Nk2Rk} we can determine the reliability coefficients for the two-terminal reliability rule, by analyzing the four motifs as shown in Figure~\ref{fig:subgraph3}. Results of this computation is shown in Table~\ref{tab:toynettale}.\par
For two dimensional grid, we study only the left case of two cases in Figure~\ref{fig:2dgrids}. For two terminal reliability from $S$ to $T$ there exist 184 structural motifs of sizes 6 to 14, thus it is not trivial to draw all structural motifs and their overlap. We provide computation for $N_k$ and $R_k$ for $k \le 10 $ in Table~\ref{tab:2d_grid}.
\color{black}
\subsection{Reliability in terms of structural motifs}
\color{black}
Given the rather complicated relationship between $R_k$ and $N_k$ in Equation~\ref{eq:Nk2Rk}, it is somewhat surprising that $R(x)$ can be expressed very simply in terms of $N_k$.
Consider the contribution of a single structural motif of size $k_0$ to $R(x)$.
Using Equation~\ref{eq:Nk2Rk}, $R_k = {E-k_0 \choose k-k_0}$.
This set of coefficients determines $R(x)$
\begin{eqnarray}
R(x) &\equiv& \sum_{k=0}^E R_k x^k (1-x)^{E-k} \nonumber \\
   &=& \sum_{k=0}^E {E-k_0 \choose k-k_0} x^k (1-x)^{E-k} \nonumber  \\
    &=& x^{k_0} \sum_{k=k_0}^E {E-k_0 \choose k-k_0} x^{k-k_0} (1-x)^{E-k_0 - (k-k_0)} \nonumber \\
    &=& x^{k_0} \sum_{k'=0}^{E-k_0} {E-k_0 \choose k'} x^{k'} (1-x)^{E-k_0 - k'} \nonumber  \\
    &=& x^{k_0}
\end{eqnarray}
Since the effect of each structural motif, and each motif overlap, is additive on $R_k$, we can reduce the general case to sums like the above, so we immediately find:
\begin{equation}
\label{eq:RxNk}
R(x) = \sum_{k=0}^E N_k x^k
\end{equation}
Thus the $N_k$ defined in Equation~\ref{eq:Nk} are indeed the same coefficients as those introduced in Equation~\ref{eq:Ndef}.
\begin{table}
\caption{\label{tab:toynettale} By inspection, we have the values for $N^{(l)}_k$ given in the table on the right, and the values of $N_k$ and $R_k$ as given by Equations~\ref{eq:Nk} and \ref{eq:Nk2Rk}.} 
\begin{tabular}{|c|c|c|c|c|}
\hline
$l$ & $k$ & $N^{(l)}_k$ & $N_k$ & $R_k$\\
\hline
1 & 3 & 3 & 3 & 3\\
\hline
   & 4 &     & 0 & 12 \\ 
 \hline
1 & 5 & 1 & & \\
2 & 5 & 2 & -1 &17 \\
\hline
2 & 6 & 3 & -3 & 7\\
\hline
2 & 7 & 1 & & \\
3 & 7 & 4 & & \\
4 & 7 & 1 & 2& 1\\
\hline
\end{tabular}
\end{table}

\begin{table}
\caption{\label{tab:2d_grid} Number of structural motifs and their overlap for the two terminal reliability rule on the left grid in Figure
\ref{fig:2dgrids}.}
\begin{center}
\begin{tabular}{|m{0.48cm}|m{0.58cm}|m{0.68cm}|m{0.58cm}|m{0.58cm}|m{0.82cm}|@{}m{0pt}@{}}
\hline
$k$ & $N_{k}^{1}$ & $N_{k}^{2}$ & $N_{k}^{3}$ & $N_k $ & $R_k$&\\[2ex]
\hline
5 & 0 & 0 & 0 &0 & 0\\ 
\hline
6 & 20 & 0 & 0 &20 & 20&\\[4pt]
\hline
7 & 0 & 0 & 0 & 0 & 360&\\[4pt] 
\hline
8 & 36 & -30 & 0 & 6 & 3066 &\\[4pt] 
\hline
9 & 0 & -84 & 0 & -84 & 16332 &\\[4pt] 
\hline
10 & 48 & -146 & 144 & 10 & 60670 &\\[4pt] 
\hline
\end{tabular}
\end{center}
\end{table}

\color{black}

\subsection{Alternative damage models}

The reasoning above is all done in the context of the usual edge damage model introduced by Moore and Shannon.
This damage model is appropriate for studying bond percolation.
An entirely analogous set of arguments applies to a vertex damage model, in which a set of $k$ {\em vertices} is chosen uniformly at random, producing a unique subgraph containing all the edges whose endpoints are both in the selected set of vertices. 
This damage model is appropriate for studying site percolation.
Coefficients analogous to $P_k$ and $N_k$ can be derived (substituting the number of vertices $V$ for the number of edges $E$ wherever it appears) and structural motifs can be defined in terms of vertex removal instead of edge removal. The physical interpretation of $N_k$ in terms of these structural motifs is the same. 
It is likely that there are many other damage models with these properties. Here, we consider only the edge damage model, because it serves to illustrate the role of structural motifs and its analysis is simpler.

\section{Estimates and bounds on reliability for special cases}

In this section, we illustrate how the study of structural motifs and their overlaps helps understand network reliability under several different rules.

\subsection{Exact expressions}
If the network contains $m$ structural motifs that are disjoint, and they all have the same size $k_0$, the only non-zero coefficients are $N_{l k_0}^{(l)} = (-1)^{l+1} {m \choose l}$,
yielding 
\begin{equation}
R(x) = 1 - (1-x^{k_0})^m.
\end{equation}
If the network contains $m$ structural motifs, every pair overlaps in all but one edge, and they all have the same size $k_0$, the only non-zero coefficients are
$N_{k_0+l-1}^{(l)} = (-1)^{l+1}{m \choose l}$,
yielding 
\begin{equation}
R(x) = x^{k_0-1} [1-(1-x)^m].
\end{equation}

We can use the identities given in Eq.~\ref{eq:NkSum} and \ref{eq:AbsNkSum} to evaluate $N_k$ for the following case:
there are $f$ structural motifs;
they all have the same size $k_0$;
and all unions of structural motifs have one of only one or two other sizes.
Although this case is somewhat artificial, note that the first two conditions are satisfied for any network under the $AR$-$\alpha$ rule.
It seems likely that the last restriction can be relaxed if additional combinatorial identities are brought to bear on the problem.
First suppose that the only nonzero coefficients $N_k$ are for $k_0$ and $k_1$. Then we must have the following:
\begin{eqnarray}
N_{k_0} &=& m; \nonumber \\
m + N_{k_1} &=& 1; \\
m + |N_{k_1}| &=& 2^m - 1. \nonumber
\end{eqnarray}
These simultaneous equations admit a solution only for $m=2$, for which $N_{k_1} = 1-m$ and hence
\begin{equation}
R(x) = x^{k_0}\left[2 - x^{k_1 - k_0} \right].
\end{equation}
Note that $k_1$ is not determined by this argument; however, it is easy to see that $k_0+1 \le k_1 \le mk_0$.
Now consider the above case, but with three nonzero coefficients instead of two. We have:
\begin{eqnarray}
N_{k_0} &=& m; \nonumber \\
m + N_{k_1} + N_{k_2} &=& 1; \\
m + |N_{k_1}| + |N_{k_2}| &=& 2^m - 1. \nonumber
\end{eqnarray}
If we look for solutions with $N_{k_2} \ge 1$, we must have $m \ge 3$.
Then the solution is 
\begin{equation}
N_{k_1}= 1-2^{m-1}; \qquad N_{k_2} = 2^{m-1} - m.
\end{equation}

This gives 
\begin{equation}
R(x) = m(x^{k_0} - x^{k_2}) + x^{k_1} + 2^{m-1}(x^{k_2}-x^{k_1}).
\end{equation}

\subsection{Perturbative estimates of reliability}
Since $R(x)$ is defined for $x$ in the interval $[0,1]$,
it is tempting to think that the lowest-order term in $x$ that appears in the reliability polynomial, i.e., $N_{k_{min}} x^{k_{min}}$, is a good estimate of its value.
Note that $N_{k_{min}} = N^{(1)}_{k_{min}}$, since any union of 2 or more structural motifs must contain more than $k_{min}$ edges.
Moreover $N_{k_{min}} = R_{k_{min}}$.
Unfortunately, because the coefficients $N_k$ may grow combinatorially and may be either positive or negative, the leading order coefficient may not be sufficient to determine behavior of the reliability polynomial far from 0.
For example, if there are very few different structural motifs with $k_{min}$ edges (specifically, if $N_{k_{min}} \ll N_{1+k_{min}} x$), the contribution of $N_{k_{min}}$ may be overwhelmed by larger structural motifs.
Nevertheless, evaluating the lowest-order term provides insight into the relationship between graph structure and reliability. 


\begin{itemize}[wide]

\item{\bf All-Terminal reliability}:
Recall that the structural motifs for the All-Terminal rule are spanning trees.
Each such tree has exactly $V-1$ edges.
$N_{V-1}$ is thus the number of spanning trees, so
 the lowest-order term in the reliability polynomial is $N_{V-1}x^{V-1}$.
The (Kirchhoff) Matrix Tree Theorem~\cite{Kirchhoff:1847} gives $N_{V-1}$ in terms of a cofactor of the graph Laplacian matrix.


\item{\bf $AR$-$\alpha$ reliability}:
The structural motifs for the $AR$-$\alpha$ reliability rule are trees that contain at least $\alpha V$ vertices.
Letting $t$ be the number of such trees, the leading order term in $R(x)$ is $t x^{\alpha V-1}$.
Higher-order terms depend on how the trees overlap. 
We can use this to establish a tight lower bound on $R(x)$ for
one particular choice of $\alpha$.

The lower bound is generated by graphs that minimize the coefficient of the next higher order term $x^{\alpha V}$.
This in turn requires that as many as possible of the motifs overlap in all but one edge.
For example, beginning with a single tree, we can change one edge to any other edge that is not already in the tree and does not create a loop in the tree.
There are at most $E - (\alpha V-1)$ ways to do this, depending on the graph.
Thus there is a graph with $t$ trees, each of which contains $\alpha$ vertices, each of which differs from any other by exactly 2 edges, if and only if $t \le E+2-\alpha V$.
In this case, $R(x) = x^{\alpha V-2}(1-x)^t$.
As far as we know, this particular tree structure occurs only for $\alpha V=E-1$.
The graph in which it occurs has a central vertex of degree $t$ connected to $t$ linear chains of length $E/t$ (thus $t$ must divide $E$ evenly).
The trees contain every edge except the last edge on one of the chains.


\item{\bf $EAR$-$\alpha$ reliability}:
Satisfying the $EAR$-$\alpha$ rule demands that the sum of squared component sizes equals or exceeds $\alpha V^2$.
What are the structural motifs for this rule?
Consider a partition $\Pi$ of $V$, i.e., a set of positive integers $\pi_i$ whose sum is $V$.
The number of elements in $\Pi$ varies from one partition to another.
Then $\pi_i$ could represent the number of vertices in the $i$th connected component.
Furthermore, if each component is a tree, the number of edges in the $i$th component is just $\pi_i-1$, hence the number of edges in the entire subgraph is $\sum_i (\pi_i - 1) = V - C$, where $C = |\Pi|$ is the number of components.
There are many ways to assign vertices to components, even for a single $\Pi$.
Each will generate a different structural motif, as long as the reliability condition
$\sum_i \pi_i^2 \ge \alpha V^2$ is satisfied.
The smallest number of edges results from a subgraph with the largest number of components.
The result is that $k_{min}$, the size of the smallest structural motif, is the size of a subgraph with all isolated vertices except for one large tree with $v$ vertices.
$k_{min}$ can be determined by the constraint
\begin{equation}
\sum_i \pi_i^2  = v^2 + (V-v) \ge \alpha V^2,
\end{equation}
or
\begin{equation}
v > \sqrt{\alpha} \left[1 - \frac{1}{\alpha V}(1 - \frac{1}{4 V})\right]^{-1/2} V.
\end{equation}
Thus $k_{min} = V - (V - v + 1) \approx \sqrt{\alpha} V - 1$, and $N_{k_{min}}$ is the number of different trees that can be made with $k_{min}$ edges.
\end{itemize}

\section{Structural motifs to find edge importance}

\begin{figure}[htbp]
\includegraphics[width=3in]{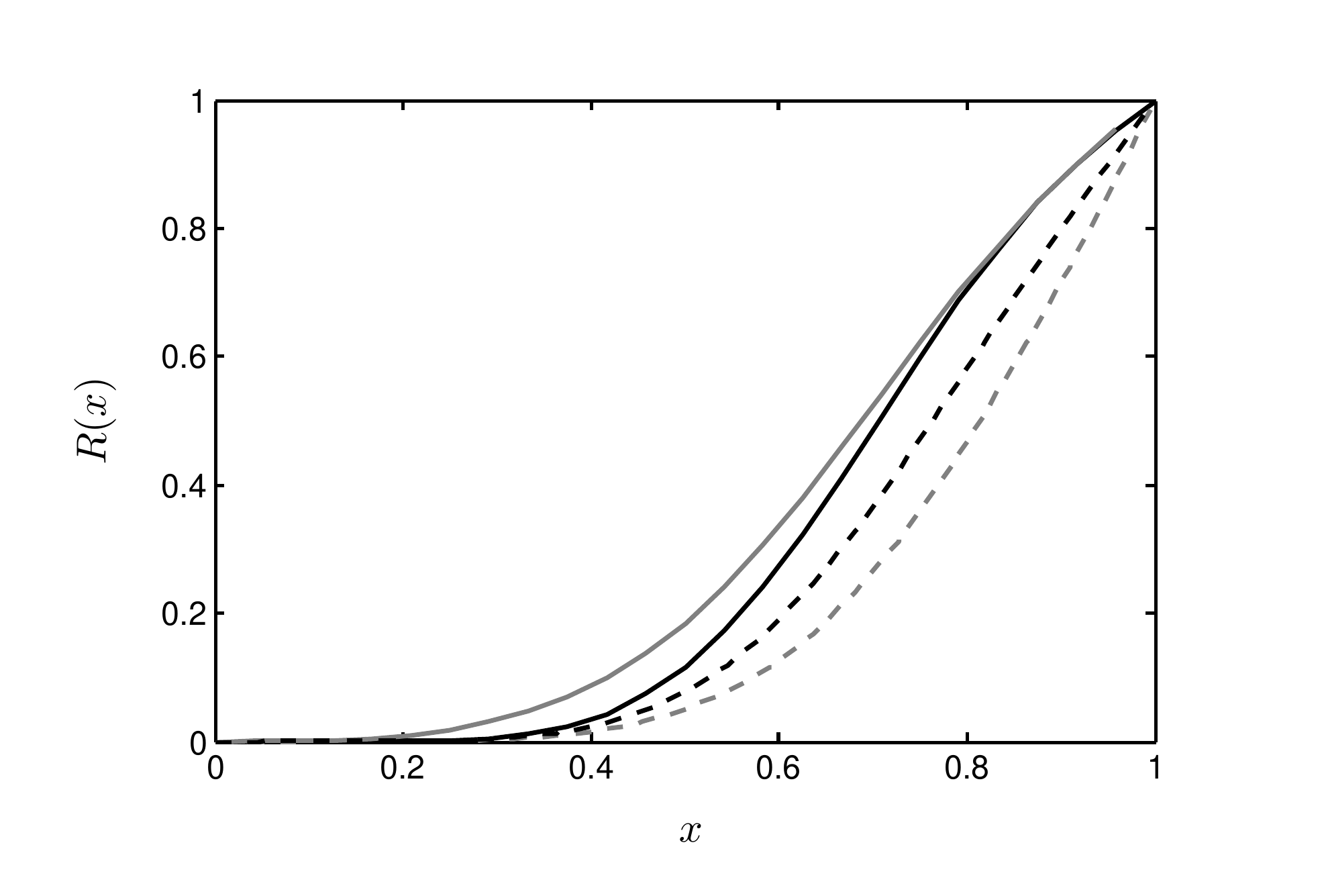}
\caption{Reliability polynomial for two terminal reliability black curve represents reliability for $S-T_{1}$, gray curve for $S-T_{2}$ and dashed curves represent the reliability polynomial after removing two edges. } 
\label{fig:rel_grid}
\end{figure}

In section ~\ref{sec:st_motifs} we demonstrated how structural motifs for 2 dimensional grid depend on parameters of the two terminal reliability rule i.e. $S$ and $T$. Here we explain how this fact can effect edge importance based on reliability rules. We computed reliability for two-terminal reliability rule for two case on two dimensional grids in Figure~\ref{fig:2dgrids}. It is clear that grid is more reliable for the reliability rule parameters in the right\i.e. it is more probable to have a path from $S$ to $T_2$ than to $T_1$. This can be seen in Figure~\ref{fig:rel_grid}.
Next we remove two out of three edges on the shortest path connecting $S$ and $T_2$ and we compute the reliability for both cases again. We see that reliability of the grid decreases more for the left case as expected. This result suggests to employ structural motifs for finding most important edges in a way that reflects the choices of parameters for the reliability rule.

\begin{figure}[htbp]
\includegraphics[width=2in]{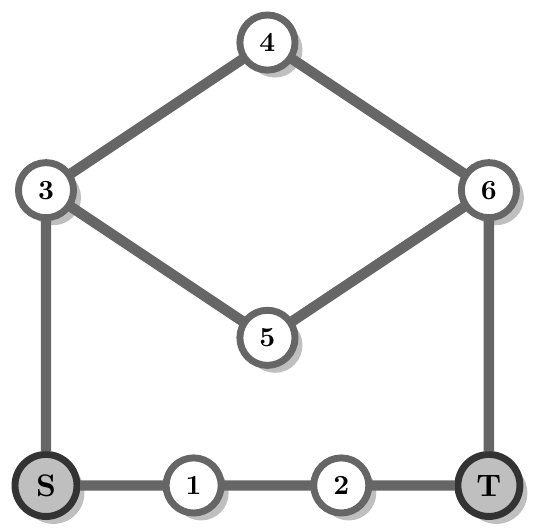}
\caption{Toy graph illustrating the possible $x$-dependence of an edge's importance.}
\label{fig:toy2}
\end{figure}

The reliability polynomials of different graphs may intersect multiple times.~\cite{crossingreliability1}
This means that, for different values of the damage model's parameters $\vec{x}$, the relative reliability of the two graphs switches signs.
If both graphs are subgraphs of the network of interest with the same number of elements removed, then clearly the Birnbaum importance ranking of the elements is different for different parameter values.
This is perhaps surprising, but it is an important feature of this approach compared to, for example, a graph statistic that is independent of $\vec{x}$.

For example, consider the graph in Figure~\ref{fig:toy2} under $S-T$ reliability for the indicated $S$ and $T$.
First, we write down the reliability for the graph by inspection from its structural motifs.
There are three motifs, $A\equiv S12T$, $B\equiv S354T$, and $C\equiv S364T$. The first has size 3; the second and third, size $4$. The second and third overlap in two edges, but are disjoint from the first. 
There are no edges that do not appear in any structural motif.
Taken together, this gives:
\begin{equation}
R(x) = x^3 + 2x^4 - x^6 - 2x^7 + x^9.
\end{equation}
(Note that $R(x)$ satisfies the constraints $\sum N_k = 1$ and $\sum |N_k| = 2^m-1$, where $m=3$ is the number of structural motifs.)
By symmetry, we expect the three edges $S1$, $12$, and $2T$ to be equally important, and also the pair $S3$ and $4T$, and finally the four edges $35$, $36$, $54$, and $64$.
Which edge is most important? A moment's thought shows that any edge from the last four is less important than any other edge.
The real choice is between $S1$, which is part of a single structural motif of size 3, and $S3$, which is part of {\em two} structural motifs of size 4.
We consider the reliability $R_1(x)$ of the graph after removing edge $S1$, leaving motifs $B$ and $C$, and the reliability $R_2(x)$ after removing edge $S3$, leaving only motif $A$.
Again by inspection, these are:
\begin{eqnarray}
R_1(x) &=& 2x^4 - x^6\\
R_2(x) &=& x^3.
\end{eqnarray}
By definition, the importance of the edges is $I_{S1}(x) \equiv R(x) - R_1(x)$ and $I_{S2}(x) \equiv R(x) - R_2(x)$.
Hence, the rank of the edges switches if the polynomial $I_{S1}(x) - I_{S2}(x) = R_2(x) - R_1(x)$ changes sign.
In fact, this polynomial has a zero in the interval $[0,1]$.
That is,
\begin{equation}
R_2(x) - R_1(x) = x^3(1 - 2x + x^3) \left\{
\begin{array}{cl}
>0 & {\rm for\ }x< 0.618... \\
<0 &{\rm for\ } 0.618... < x
\end{array}
\right.
\end{equation}

\section{Applying reliability concepts to other network analysis problems}

\begin{figure}
\begin{center}
\includegraphics[width=3.0in]{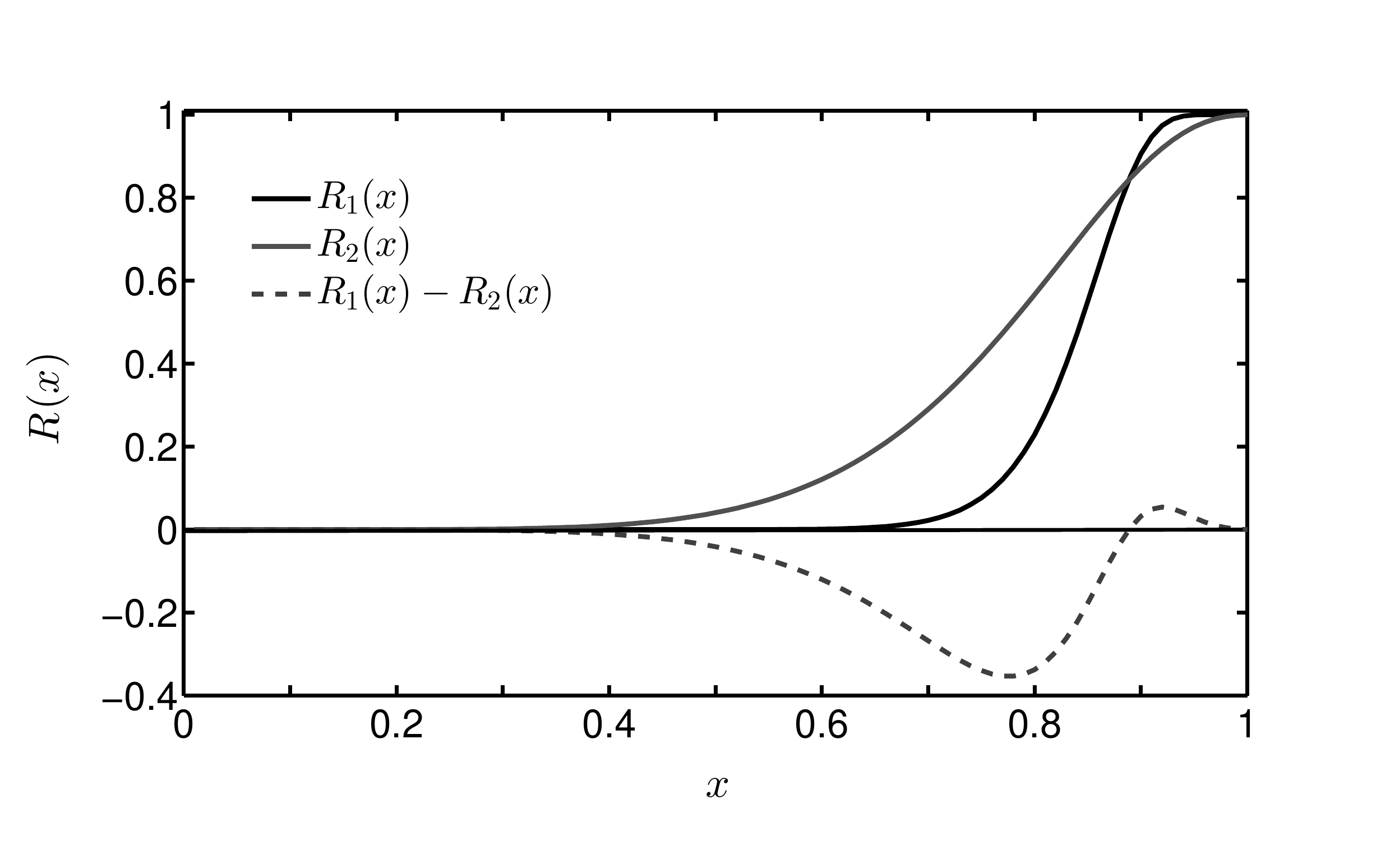}
\caption{Comparing the reliability of a network with many overlapping structural motifs with that of a network with a few disjoint motifs.}
\label{fig:TradeOffs}
\end{center}
\end{figure}

The representation of the reliability polynomial in terms of structural motifs provides a convenient organizing principle for thinking about general network analysis problems. As one example, consider the tradeoffs between two systems: one with only a few completely redundant reliable subsytems and another with more, but only partially redundant, ones. To study this we consider two extreme cases of overlap. One contains $r_1$ structural motifs of size $k_1$, any two of which differ by only two edges. They are thus built using a total of $2r_1 + k_1 - 2$ edges. The reliability of this combination can be written as:
\begin{eqnarray}
R_1(x) = \sum_{i=1}^{r_1}(-1)^{i+1}\binom{r_1}{i}x^{k_1+r_1(i-1)}.
\end{eqnarray}
Using the same number of edges we can construct $r_2=\frac{2r_1+k_1-2}{k_2}$ motifs of size $k_2$ that are completely disjoint. The reliability of this combination of motifs is:
\begin{eqnarray}
R_2(x) = \sum_{i=1}^{r_2}(-1)^{i+1}\binom{r_2}{i}x^{ik_2}.
\end{eqnarray}
Knowing the reliability for these two cases, we are able to compare the reliability of networks with different configurations of structural motifs of different sizes. As an example we compared the reliability of a network composed of 20 motifs with 18 edges that are different from one another only in two edges with a network of 4 completely disjoint motifs of size 6. Figure~\ref{fig:TradeOffs} shows the reliability curves for these two networks and their difference as a function of $x$. The analysis shows that the network of disjoint motifs is more reliable for smaller values of $x$ while the opposite is true for larger $x$ values.

This approach could also be used to estimate the number of spanning trees in a graph. A spanning tree is a subgraph of the network that includes all vertices \cite{SpanTree1,SpanTree2,SpanTree3}; the number of spanning trees can be estimated by evaluating the All-Terminal reliability.
Another problem that can be addressed using this method is to identify chordless loops of various sizes in a network. A chordless loop is a sequence of vertices with more than three vertices if for all $i=1,\cdots ,k$  there is exactly one link from vertex $v_i$ to $v_{i+1}$ and there is no other link between any two of the vertices in this sequence \cite{ComplexSciences}. Recent studies on ecological networks have discovered the existence of many chordless cycles in these networks \cite{FoodWeb}, therefore enumeration of all chordless cycles can make a significant impact on understanding the structure of these networks. An appropriately-designed reliability rule can be used to count the number of chordless cycles of different sizes.\par

\section{Conclusion and future work}
In this paper we focused on the representation of the reliability polynomial in terms of structural motifs.  We have shown that network reliability is simply related to the number of edges in unions of structural motifs $N_{k}$ (\ref{eq:RxNk}). Whereas the coefficients $P_k$ of $x^k(1-x)^{E-k}$ are easy to estimate numerically but hard to work with analytically, the coefficients $N_{k}$ of $x^k$ are hard to estimate numerically but easy to work with analytically. To demonstrate this, we have derived closed-form expressions for $N_{k}$  for several types of graphs. The resulting expressions were confirmed by numerical estimation. We anticipate that this approach can lead us to a measure of edge centrality that relates the importance of an edge to the frequency of its appearance in different structural motifs \cite{edgerank}. 
While we can use numerical simulation to study specific large, realistic networks -- including epidemiology on social networks \cite{Eubank:04,SocialNet,MixPattern} -- we can use the notion of structural motifs to understand the differences between networks that are discovered in simulation.
We expect this approach to be particularly useful in studying the stability and robustness of interconnected networks \cite{InterNet,Robustness,robustness1,robustness2,Youssef:11b,robustness4}.
\color{black}

\begin{acknowledgments}
This research was partially supported by NSF NetSE Grant CNS-1011769, DTRA R\&D Grant HDTRA1-0901-0017, and DTRA CNIMS Grant HDTRA1-07-C-0113. The work described in this paper was funded by the National Institute of General Medical Sciences of the National Institutes of Health under NIH MIDAS Grant 2U01GM070694-09. We would like to acknowledge many useful comments from our external collaborators and members of the Network Dynamics and Simulation Science Laboratory (NDSSL)  particularly M. Marathe and A. Vullikanti. The content is solely the responsibility of the authors and does not necessarily represent the official views of the National Institutes of Health or DTRA.


\end{acknowledgments}

\appendix

\section{\label{sec:ExcIncProog} Inclusion-Exclusion proof of reliability polynomial}

We begin by rewriting the expansion
\begin{align}\label{first}
R (x) = \sum_{k=0}^E\; R_k\, x^k\, (1-x)^{E-k} \,,
\end{align}
as the following,
\begin{align}\label{second}
R (x) = \sum_{A\subseteq E}\; r(A)\, x^A\, (1-x)^{E-A} \,;
\end{align}
here $x^A$ is short for $x^{|A|}$ where $|A|$ refers to size of the subset $A\subseteq E$ that characterizes the edge-induced subgraph $A$. $(1-x)^{E-A}$, then,
has a similar interpretation. We note that \eqn{first} is equivalent to \eqn{second} simply because $R_k$ is counting the number of edge-induced subgraphs
$A\subseteq E$ of size $k$ that satisfy the reliability rule $r$; for these subgraphs $r(A)=1$ and the equivalence follows --for those that do not, $r(A)=0$.

If the reliability rule is coherent, there is a unique family of minimal subgraphs $A\subseteq E$ such that for every subgraph $A^\prime$ that is a \emph{proper}
subgraph of either one of them, $r(A^\prime)=0$. This means that every subgraph that is accepted by the reliability rule $r$, contains at least one of these minimal
subgraphs (entirely). Therefore, we can see that every reliability rule is in 1-1 correspondence with a certain family of (edge-induced) subgraphs $A$ of $E$. As a
result, we can now define the reliability rule in terms of the very family of \emph{minimal} subgraphs that were obtained from this rule,
\begin{align}
r(B)=\left\{\begin{array}{lcr} 1 &;& \exists\; A\in \mathcal{A}\; : \; A\subseteq B \\0 &;& \text{otherwise} \end{array} \right. \,;
\end{align}
here $\mathcal{A}$ represents this family.

We now show that $R(x)$ stands for the probability that at least one of these minimal subgraphs is \emph{operating} in the sense that all of its corresponding edges
function. Put differently, we show that
\begin{align}\label{third}
R(x) &\overset{?}{=}\; \mathbf{Pr}\; \left(\;\bigvee_i\; \{A_i \text{ operates} \}\right) \notag\\
&=\; \mathbf{Pr}\; \left(\; \bigvee_i\; A_i \right) \,,
\end{align}
where $A_i$'s are different members of the above family of minimal subgraphs indexed by $i$; the second line is understood as a shorthand notation for the first.

Let $\ang{A_1 \vee A_2}$ stand for the family of all subgraphs that contain all of the edges in at least one of the subgraphs $A_1$ or $A_2$. Also let
$\ang{\overline{A_1 \vee A_2}}$ denote the family of all subgraphs that lack at least one edge from each of the subgraphs $A_1$ and $A_2$. Next, we define
$\ang{A_1 \wedge A_2}$ as the family of all subgraphs that contain all of the edges in both $A_1$ and $A_2$. And finally, let $\ang{A_1 \wedge \overline{A_2}}$
represent the family of all subgraphs that contain all of the edges in subgraph $A_1$ \emph{and} lack at least one edge from $A_2$. We also utilize unambiguous
generalizations of these notations at a later inductive step.

Now suppose that the reliability rule $r$ is such that only one minimal subgraph, namely $A_1$, is accepted and, as a result, $R(x)$ becomes
\begin{align}
R(x) &=\; \sum_{A_1 \subseteq A \subseteq E}\; x^A\,(1-x)^{E-A} \notag\\
&=\; \sum_{A \in \ang{A_1}}\; x^A\, (1-x)^{E-A} \notag\\
&=\; x^{A_1} \label{shortened}\\
&=\; \mathbf{Pr}\; \Bigl( A_1 \text{ operates} \Bigr)  \label{single} \,;
\end{align}
where equation \eqn{shortened} can be easily obtained by induction on the edges in subgraph $E-A_1$. We can take \eqn{single} as the base case for an inductive
proof of the equivalence of expressions \eqn{second} and \eqn{third} and proceed to the induction step. However, to break in the notation we introduced above, we
also prove the case of two minimal subgraphs, and then proceed to the inductive step. Therefore, suppose that we have two minimal subgraphs
$A_1$ and $A_2$ and we want to prove the following,
\begin{align}
\mathbf{Pr}\; \left(\; \bigvee_{i\le 2}\; A_i \right) \overset{?}{=} \sum_{A\in\ang{A_1 \vee A_2}}\; x^A\, (1-x)^{E-A} \,.
\end{align}
The right hand side of this euqality is precisely the expansion of $R(x)$ for a reliability rule that is in 1-1 correspondence with the family of two
minimal subgraphs $A_1$ and $A_2$; therefore our claim is proved for this case if the above equality holds. We start from the left hand side and note that
\begin{align}\label{in-ex}
\mathbf{Pr}\; \left(\; \bigvee_{i\le 2}\; A_i \right) = \mathbf{Pr}\;\left(A_1\right) + \mathbf{Pr}\;\left(A_2\right) - \mathbf{Pr}\;\left(A_1 \wedge A_2\right) \,.
\end{align}
Now, we make the following observations,
\begin{align}
\mathbf{Pr}\;\left(A_1\right) &=\; \sum_{A\in\ang{A_1}}\; x^A\, (1-x)^{E-A} \notag\\
&=\; \left(\sum_{A\in\ang{A_1 \wedge \overline{A_2}}} + \sum_{A\in\ang{A_1 \wedge A_2}}\right)\; x^A\, (1-x)^{E-A} \,, \label{squint}\\
\mathbf{Pr}\;\left(A_1 \wedge A_2 \right) &=\; \sum_{A\in\ang{A_1 \wedge A_2}}\; x^A\, (1-x)^{E-A} \,.\label{squintt}
\end{align}
From these expansions, it immediately follows for \eqn{in-ex} that
\begin{align}
&\mathbf{Pr}\; \left(\; \bigvee_{i\le 2}\; A_i \right) \notag\\
&\hspace{2.75em}=\; \left(\sum_{A\in\ang{A_1 \wedge \overline{A_2}}} + \sum_{A\in\ang{\overline{A_1} \wedge A_2 }} + \sum_{A\in\ang{A_1 \wedge A_2}}\right)\; x^A\,
  (1-x)^{E-A} \notag\\
&\hspace{2.75em}=\; \sum_{A\in\ang{A_1 \vee A_2}}\; x^A\, (1-x)^{E-A} \,.
\end{align}
This proves the base case of a family with only two minimal subgraphs $A_1$ and $A_2$. The inductive step is quite similar,
\begin{align}
&\mathbf{Pr}\; \left(\; \bigvee_{i\le n}\; A_i \right) \notag\\
&\hspace{0.0em}=\; \mathbf{Pr}\;\left(\bigvee_{i<n} A_i\right) + \mathbf{Pr}\;\left(A_n\right) - \mathbf{Pr}\;\left(\left(\bigvee_{i<n} A_i \right)
  \wedge A_n\right) \notag\\
&\hspace{0.0em}=\; \left(\sum_{A\in\ang{\bigvee_{i<n} A_i\, \wedge\, \overline{A_n}}} + \sum_{A\in\ang{\overline{\bigvee_{i<n} A_i}\, \wedge\, A_n}} \right. \\ 
&\hspace{0.0em}+\;  \left. \sum_{A\in\ang{\bigvee_{i<n} A_i\, \wedge\, A_n}}\right)\; x^A\,
  (1-x)^{E-A} \label{explained}\\
&\hspace{0.0em}=\; \sum_{A\in\ang{\bigvee_{i<n} A_i}}\; x^A\, (1-x)^{E-A} \label{simplified}\\
&\hspace{0.0em}=\; \sum_{A\subseteq E}\; r(A)\, x^A\, (1-x)^{E-A} \,, \notag
\end{align}
where $r$ is the reliability rule that corresponds to the family of minimal subgraphs $\{A_1,\cdots,A_n\}$. In \eqn{explained},
$\ang{\,\bigvee_{i<n} A_i\, \wedge\, \overline{A_n}\,}$ represents the family of all subgraphs that contain all of the edges in at least one of the subgraphs $A_1$
through $A_{n-1}$ \emph{and} lack at least one edge from the subgraph $A_n$. Next, $\ang{\,\overline{\bigvee_{i<n} A_i}\, \wedge\, A_n\, }$ stands for the family of
all subgraphs that contain all of the edges in subgraph $A_n$ \emph{and} lack at least one edge from each and every subgraph $A_1$ through $A_{n-1}$. Lastly,
$\ang{\,\bigvee_{i<n} A_i\, \wedge\, A_n\, }$ refers to the family of all subgraphs that contain all of the edges in both $A_n$ and at least on of the subgraphs $A_1$
through $A_{n-1}$. Equation \eqn{simplified}, then, follows from the fact that the combination of these three possibilities is precisely what $\ang{\,\bigvee_{i<n}
  A_i\,}$ stands for.

So far, we have established that
\begin{align}\label{third+}
R(x) =\; \mathbf{Pr}\; \left(\; \bigvee_i\; A_i \right) \,,
\end{align}
and now we proceed to prove a final equivalent expansion, namely
\begin{align}\label{forth}
R(x) &\overset{?}{=}\; \sum_{k=0}^E\; N_k\, x^k \\
N_k &\equiv\; \sum_{l=1}^E (-1)^{l+1}\,N_k^{(l)} \,,
\end{align}
where $N_k^{(l)}$ denotes the number of combinations of $l$ minimal subgraphs whose union contains exactly $k$ edges. To show that the above holds, we appeal to equation \eqn{third+} which can be now, by account of inclusion-exclusion principle, expanded as
\begin{align}
&\mathbf{Pr}\; \left(\; \bigvee_{1\le i\le n}\; A_i \right) \notag\\
&\hspace{3.25em}=\; \sum_{i=1}^n\; (-1)^{i+1}\; \sum_{1\le j_1 < \cdots <j_i \le n}\; \mathbf{Pr}\; \left(A_{j_1} \wedge \cdots \wedge A_{j_i}\right) \notag\\
&\hspace{3.25em}=\; \sum_{i=1}^n\; (-1)^{i+1}\; \sum_{1\le j_1 < \cdots <j_i \le n}\; x^{A_{j_1} \cup \cdots \cup A_{j_i}} \,;
\end{align}
this clearly coincides with the expansion \eqn{forth}.

\section{\label{sec:ExcIncExamples} Additional examples of contribution of structural motifs to $R(x)$ }

\subsubsection{\label{sec:2motifs}Example 2: Two disjoint structural motifs}
Suppose the network $G$ contains exactly two structural motifs, that both have $k_0$ edges, and that no edge is in both.
Arguing as in Example~1, for $k<2k_0$, $R_k$ is simply twice what it is for the case of a single structural motif.
But when $k = 2k_0$, the subgraph that consists of the union of the two structural motifs will have been counted twice instead of once.
Similarly, for $k>2k_0$, the number of graphs overcounted is given by assigning $2k_0$ of the edges and choosing the remaining $k-2k_0$ in the subgraph from among the remaining $E-2k_0$ in the graph $G$.
Hence:
\begin{equation}
R_k = \left\{
\begin{array}{lr}
0 & k < k_0 \\
2{E-k_0 \choose k-k_0} & k_0 \le k < 2k_0 \\
2{E-k_0 \choose k-k_0} - {E-2k_0 \choose k-2k_0} & 2k_0 \le k
\end{array}
\right.
\end{equation}


\subsubsection{\label{sec:3motifs}Example 3: Three disjoint structural motifs}
Suppose the network contains exactly three structural motifs, that all three have $k_0$ edges, and that the three edge sets are disjoint.
Again, when $k_0\le k<2k_0$, each motif generates ${E-k_0 \choose k-k_0}$ different reliable subgraphs, and
for $2k_0 = k$, three of these subgraphs are counted twice.
But in this case, when $k$ reaches $3k_0$, the subgraph consisting of all three motifs is first included three times (once for each motif), then excluded three times (once for each pair of motifs) with the net result that it must be included again:
\begin{equation}
R_k = \left\{
\begin{array}{lr}
0 & k < k_0 \\
3{E-k_0 \choose k-k_0} & k_0 \le k < 2k_0\\
3{E-k_0 \choose k-k_0} - 3{E-2k_0 \choose k-2k_0} & 2k_0 \le k < 3k_0 \\
3{E-k_0 \choose k-k_0} - 3{E-2k_0 \choose k-2k_0} + {E-3k_0 \choose k-3k_0} & 3k_0 \le k
\end{array}
\right.
\end{equation}
\section{\label{sec:Nkconstraints} Constraints on coefficients}

Several constraints apply to $N_k$. A union of $l$ motifs can have size $k$ only if all possible unions of $l-1$ of the same motifs have size less than $k$.
This leads to a set of constraints of the form
\begin{equation}
N^{(2)}_k \le {\sum_{k'=0}^k N^{(1)}_{k'} \choose 2}.
\end{equation}
In addition, a union of $l$ motifs can have size $k$ only if all possible unions of $l-1$ and $l-2$ of the same motifs have size less than $k$. For instance, $N_{k}^{(3)}$ has the following upper bound
\begin{equation}
N_{k}^{(3)} \leq (\sum_{k'=0}^{k'=k} N_{k'}^{(1)})(\sum_{k'=0}^{k'=k} N_{k'}^{(2)}) + \binom{\sum_{k'=0}^{k'=k} N_{k'}^{(1)}}{3}.
\end{equation}

Overall, since all unions of $l$ structural motifs must be included in $G$, we have
\begin{equation}
\sum_{k=0}^E  N^{(l)}_k  = {f \choose l},
\label{eq:NklSum}
\end{equation}
where $f$ is the total number of structural motifs.
Finally, the facts that $\sum_{l=0}^{f} {f \choose l} = 2^f$ and $\sum_{l=0}^{f} (-1)^l {f \choose l} = 0$ imply that
\begin{eqnarray}
\sum_{k=0}^E N_k  &=& \sum_{k=0}^E \sum_{l=1}^f (-1)^{l+1} N^{(l)}_k \nonumber \\
&=& \sum_{l=1}^f (-1)^{l+1} {f \choose l} = 1 ;
\label{eq:NkSum}
\end{eqnarray}
and 
\begin{eqnarray}
\sum_{k=0}^E |N_k|  &=& \sum_{k=0}^E \sum_{l=1}^f N^{(l)}_k \nonumber \\
&=& \sum_{l=1}^f {f \choose l} = 2^f - 1 ;
\label{eq:AbsNkSum}
\end{eqnarray}

\bibliography{NKpaper}

\begin{thebibliography}{27}%
\makeatletter
\providecommand \@ifxundefined [1]{%
 \@ifx{#1\undefined}
}%
\providecommand \@ifnum [1]{%
 \ifnum #1\expandafter \@firstoftwo
 \else \expandafter \@secondoftwo
 \fi
}%
\providecommand \@ifx [1]{%
 \ifx #1\expandafter \@firstoftwo
 \else \expandafter \@secondoftwo
 \fi
}%
\providecommand \natexlab [1]{#1}%
\providecommand \enquote  [1]{``#1''}%
\providecommand \bibnamefont  [1]{#1}%
\providecommand \bibfnamefont [1]{#1}%
\providecommand \citenamefont [1]{#1}%
\providecommand \href@noop [0]{\@secondoftwo}%
\providecommand \href [0]{\begingroup \@sanitize@url \@href}%
\providecommand \@href[1]{\@@startlink{#1}\@@href}%
\providecommand \@@href[1]{\endgroup#1\@@endlink}%
\providecommand \@sanitize@url [0]{\catcode `\\12\catcode `\$12\catcode
  `\&12\catcode `\#12\catcode `\^12\catcode `\_12\catcode `\%12\relax}%
\providecommand \@@startlink[1]{}%
\providecommand \@@endlink[0]{}%
\providecommand \url  [0]{\begingroup\@sanitize@url \@url }%
\providecommand \@url [1]{\endgroup\@href {#1}{\urlprefix }}%
\providecommand \urlprefix  [0]{URL }%
\providecommand \Eprint [0]{\href }%
\providecommand \doibase [0]{http://dx.doi.org/}%
\providecommand \selectlanguage [0]{\@gobble}%
\providecommand \bibinfo  [0]{\@secondoftwo}%
\providecommand \bibfield  [0]{\@secondoftwo}%
\providecommand \translation [1]{[#1]}%
\providecommand \BibitemOpen [0]{}%
\providecommand \bibitemStop [0]{}%
\providecommand \bibitemNoStop [0]{.\EOS\space}%
\providecommand \EOS [0]{\spacefactor3000\relax}%
\providecommand \BibitemShut  [1]{\csname bibitem#1\endcsname}%
\let\auto@bib@innerbib\@empty
\bibitem [{\citenamefont {Moore}\ and\ \citenamefont
  {Shannon}(1956)}]{Moore:56}%
  \BibitemOpen
  \bibfield  {author} {\bibinfo {author} {\bibfnamefont {E.}~\bibnamefont
  {Moore}}\ and\ \bibinfo {author} {\bibfnamefont {C.}~\bibnamefont
  {Shannon}},\ }\href@noop {} {\bibfield  {journal} {\bibinfo  {journal}
  {Journal of the Franklin Institute}\ }\textbf {\bibinfo {volume} {262}},\
  \bibinfo {pages} {191} (\bibinfo {year} {1956})}\BibitemShut {NoStop}%
\bibitem [{\citenamefont {Newman}\ \emph {et~al.}(2006)\citenamefont {Newman},
  \citenamefont {Barabási},\ and\ \citenamefont {Watts}}]{DynamicNewman}%
  \BibitemOpen
  \bibfield  {author} {\bibinfo {author} {\bibfnamefont {M.}~\bibnamefont
  {Newman}}, \bibinfo {author} {\bibfnamefont {A.}~\bibnamefont {Barabási}}, \
  and\ \bibinfo {author} {\bibfnamefont {D.}~\bibnamefont {Watts}},\
  }\href@noop {} {\bibfield  {journal} {\bibinfo  {journal} {Princeton
  University Press}\ } (\bibinfo {year} {2006})}\BibitemShut {NoStop}%
\bibitem [{\citenamefont {Ganesh}\ \emph {et~al.}(2005)\citenamefont {Ganesh},
  \citenamefont {Massoulié},\ and\ \citenamefont {Towsley}}]{Dynamic}%
  \BibitemOpen
  \bibfield  {author} {\bibinfo {author} {\bibfnamefont {A.}~\bibnamefont
  {Ganesh}}, \bibinfo {author} {\bibfnamefont {L.}~\bibnamefont {Massoulié}},
  \ and\ \bibinfo {author} {\bibfnamefont {D.}~\bibnamefont {Towsley}},\
  }\href@noop {} {\bibfield  {journal} {\bibinfo  {journal} {Proceedings of the
  IEEE INFOCOM Miami, FL, 05}\ } (\bibinfo {year} {2005})}\BibitemShut
  {NoStop}%
\bibitem [{\citenamefont {Colbourn}(1987)}]{Colbourn}%
  \BibitemOpen
  \bibfield  {author} {\bibinfo {author} {\bibfnamefont {C.~J.}\ \bibnamefont
  {Colbourn}},\ }\href@noop {} {\emph {\bibinfo {title} {The Combinatorics of
  Network Reliability}}}\ (\bibinfo  {publisher} {Oxford University Press},\
  \bibinfo {year} {1987})\BibitemShut {NoStop}%
\bibitem [{\citenamefont {Youssef}\ \emph {et~al.}(2013)\citenamefont
  {Youssef}, \citenamefont {Khorramzadeh},\ and\ \citenamefont
  {Eubank}}]{Youssef:13}%
  \BibitemOpen
  \bibfield  {author} {\bibinfo {author} {\bibfnamefont {M.}~\bibnamefont
  {Youssef}}, \bibinfo {author} {\bibfnamefont {Y.}~\bibnamefont
  {Khorramzadeh}}, \ and\ \bibinfo {author} {\bibfnamefont {S.}~\bibnamefont
  {Eubank}},\ }\href@noop {} {\bibfield  {journal} {\bibinfo  {journal}
  {Physical Review E}\ }\textbf {\bibinfo {volume} {88}} (\bibinfo {year}
  {2013})}\BibitemShut {NoStop}%
\bibitem [{\citenamefont {Jane}\ \emph {et~al.}(2009)\citenamefont {Jane},
  \citenamefont {Shenb},\ and\ \citenamefont {Laihc}}]{TwoTermCite}%
  \BibitemOpen
  \bibfield  {author} {\bibinfo {author} {\bibfnamefont {C.-C.}\ \bibnamefont
  {Jane}}, \bibinfo {author} {\bibfnamefont {W.-H.}\ \bibnamefont {Shenb}}, \
  and\ \bibinfo {author} {\bibfnamefont {Y.-W.}\ \bibnamefont {Laihc}},\
  }\href@noop {} {\bibfield  {journal} {\bibinfo  {journal} {European Journal
  of Operational Research}\ }\textbf {\bibinfo {volume} {195}},\ \bibinfo
  {pages} {427} (\bibinfo {year} {2009})}\BibitemShut {NoStop}%
\bibitem [{\citenamefont {Elmallah}(1992)}]{KTermCite}%
  \BibitemOpen
  \bibfield  {author} {\bibinfo {author} {\bibfnamefont {E.~S.}\ \bibnamefont
  {Elmallah}},\ }\href@noop {} {\bibfield  {journal} {\bibinfo  {journal}
  {Networks}\ }\textbf {\bibinfo {volume} {22}},\ \bibinfo {pages} {369}
  (\bibinfo {year} {1992})}\BibitemShut {NoStop}%
\bibitem [{\citenamefont {Rong-Hong}(1993)}]{AllTermCite}%
  \BibitemOpen
  \bibfield  {author} {\bibinfo {author} {\bibfnamefont {J.}~\bibnamefont
  {Rong-Hong}},\ }\href@noop {} {\bibfield  {journal} {\bibinfo  {journal}
  {Computers and Operations Research}\ }\textbf {\bibinfo {volume} {20}},\
  \bibinfo {pages} {25} (\bibinfo {year} {1993})}\BibitemShut {NoStop}%
\bibitem [{\citenamefont {LM}\ \emph {et~al.}(2002)\citenamefont {LM},
  \citenamefont {CP}, \citenamefont {IM},\ and\ \citenamefont
  {Simon~C}}]{PercolationEpidemiology}%
  \BibitemOpen
  \bibfield  {author} {\bibinfo {author} {\bibfnamefont {S.}~\bibnamefont
  {LM}}, \bibinfo {author} {\bibfnamefont {W.}~\bibnamefont {CP}}, \bibinfo
  {author} {\bibfnamefont {S.}~\bibnamefont {IM}}, \ and\ \bibinfo {author}
  {\bibfnamefont {K.~J.}\ \bibnamefont {Simon~C}},\ }\href@noop {} {\bibfield
  {journal} {\bibinfo  {journal} {Mathematical Biosciences 180}\ } (\bibinfo
  {year} {2002})}\BibitemShut {NoStop}%
\bibitem [{\citenamefont {Alon}(2007)}]{AlonNatRevGen}%
  \BibitemOpen
  \bibfield  {author} {\bibinfo {author} {\bibfnamefont {U.}~\bibnamefont
  {Alon}},\ }\href@noop {} {\bibfield  {journal} {\bibinfo  {journal} {Nature
  Reviews Genetics}\ }\textbf {\bibinfo {volume} {8}},\ \bibinfo {pages}
  {450,461} (\bibinfo {year} {2007})}\BibitemShut {NoStop}%
\bibitem [{\citenamefont {Kirchhoff}(1847)}]{Kirchhoff:1847}%
  \BibitemOpen
  \bibfield  {author} {\bibinfo {author} {\bibfnamefont {G.}~\bibnamefont
  {Kirchhoff}},\ }\href@noop {} {\bibfield  {journal} {\bibinfo  {journal}
  {Ann. Phys. Chem.}\ }\textbf {\bibinfo {volume} {72}},\ \bibinfo {pages}
  {497} (\bibinfo {year} {1847})}\BibitemShut {NoStop}%
\bibitem [{\citenamefont {Brown}\ \emph {et~al.}(2011)\citenamefont {Brown},
  \citenamefont {Koc},\ and\ \citenamefont {Kooij}}]{crossingreliability1}%
  \BibitemOpen
  \bibfield  {author} {\bibinfo {author} {\bibfnamefont {J.}~\bibnamefont
  {Brown}}, \bibinfo {author} {\bibfnamefont {Y.}~\bibnamefont {Koc}}, \ and\
  \bibinfo {author} {\bibfnamefont {R.}~\bibnamefont {Kooij}},\ }in\ \href@noop
  {} {\emph {\bibinfo {booktitle} {Proceedings of the 3rd International
  Congress on Ultra Modern Telecommunications and Control Systems and Workshops
  (ICUMT), 2011}}}\ (\bibinfo {address} {Budapest, Hungary, October 5-7,
  2011},\ \bibinfo {year} {2011})\BibitemShut {NoStop}%
\bibitem [{\citenamefont {Chaiken}(1982)}]{SpanTree1}%
  \BibitemOpen
  \bibfield  {author} {\bibinfo {author} {\bibfnamefont {S.}~\bibnamefont
  {Chaiken}},\ }\href@noop {} {\bibfield  {journal} {\bibinfo  {journal} {SIAM
  J. ALG. DISC. METH}\ }\textbf {\bibinfo {volume} {3}} (\bibinfo {year}
  {1982})}\BibitemShut {NoStop}%
\bibitem [{\citenamefont {Cvetković}\ \emph {et~al.}(1998)\citenamefont
  {Cvetković}, \citenamefont {Doob},\ and\ \citenamefont {Sachs}}]{SpanTree2}%
  \BibitemOpen
  \bibfield  {author} {\bibinfo {author} {\bibfnamefont {D.~M.}\ \bibnamefont
  {Cvetković}}, \bibinfo {author} {\bibfnamefont {M.}~\bibnamefont {Doob}}, \
  and\ \bibinfo {author} {\bibfnamefont {H.}~\bibnamefont {Sachs}},\
  }\href@noop {} {\emph {\bibinfo {title} {Spectra of Graphs: Theory and
  Applications}}}\ (\bibinfo  {publisher} {Wiley},\ \bibinfo {year}
  {1998})\BibitemShut {NoStop}%
\bibitem [{\citenamefont {Buekenhout}\ and\ \citenamefont
  {Parker}(1998)}]{SpanTree3}%
  \BibitemOpen
  \bibfield  {author} {\bibinfo {author} {\bibfnamefont {F.}~\bibnamefont
  {Buekenhout}}\ and\ \bibinfo {author} {\bibfnamefont {M.}~\bibnamefont
  {Parker}},\ }\href@noop {} {\bibfield  {journal} {\bibinfo  {journal} {Disc.
  Math}\ }\textbf {\bibinfo {volume} {186}},\ \bibinfo {pages} {69} (\bibinfo
  {year} {1998})}\BibitemShut {NoStop}%
\bibitem [{\citenamefont {Sokhn}\ \emph {et~al.}(2013)\citenamefont {Sokhn},
  \citenamefont {Baltensperger}, \citenamefont {Bersier}, \citenamefont
  {Hennebert},\ and\ \citenamefont {Ultes-Nitschey}}]{ComplexSciences}%
  \BibitemOpen
  \bibfield  {author} {\bibinfo {author} {\bibfnamefont {N.}~\bibnamefont
  {Sokhn}}, \bibinfo {author} {\bibfnamefont {R.}~\bibnamefont
  {Baltensperger}}, \bibinfo {author} {\bibfnamefont {L.-F.}\ \bibnamefont
  {Bersier}}, \bibinfo {author} {\bibfnamefont {J.}~\bibnamefont {Hennebert}},
  \ and\ \bibinfo {author} {\bibfnamefont {U.}~\bibnamefont {Ultes-Nitschey}},\
  }\href@noop {} {\emph {\bibinfo {title} {Identification of Chordless Cycles
  in Ecological Networks}}}\ (\bibinfo  {publisher} {Springer},\ \bibinfo
  {year} {2013})\ pp.\ \bibinfo {pages} {316--324}\BibitemShut {NoStop}%
\bibitem [{\citenamefont {Huxham}\ \emph {et~al.}(1996)\citenamefont {Huxham},
  \citenamefont {Beaney},\ and\ \citenamefont {Raffaelli}}]{FoodWeb}%
  \BibitemOpen
  \bibfield  {author} {\bibinfo {author} {\bibfnamefont {M.}~\bibnamefont
  {Huxham}}, \bibinfo {author} {\bibfnamefont {S.}~\bibnamefont {Beaney}}, \
  and\ \bibinfo {author} {\bibfnamefont {D.}~\bibnamefont {Raffaelli}},\
  }\href@noop {} {\ \textbf {\bibinfo {volume} {76}},\ \bibinfo {pages} {284}
  (\bibinfo {year} {1996})}\BibitemShut {NoStop}%
\bibitem [{\citenamefont {Eubank}\ \emph {et~al.}(2013)\citenamefont {Eubank},
  \citenamefont {Youssef},\ and\ \citenamefont {Khorramzadeh}}]{edgerank}%
  \BibitemOpen
  \bibfield  {author} {\bibinfo {author} {\bibfnamefont {S.}~\bibnamefont
  {Eubank}}, \bibinfo {author} {\bibfnamefont {M.}~\bibnamefont {Youssef}}, \
  and\ \bibinfo {author} {\bibfnamefont {Y.}~\bibnamefont {Khorramzadeh}},\
  }in\ \href@noop {} {\emph {\bibinfo {booktitle} {Proceedings of the second
  workshop on complex networks and their applications, SITIS 2013}}}\ (\bibinfo
  {address} {Tokyo, Japan},\ \bibinfo {year} {2013})\BibitemShut {NoStop}%
\bibitem [{\citenamefont {Eubank}\ \emph {et~al.}(2004)\citenamefont {Eubank},
  \citenamefont {Guclu}, \citenamefont {Kumar}, \citenamefont {Marathe},
  \citenamefont {Srinivasan}, \citenamefont {Toroczkai},\ and\ \citenamefont
  {Wang}}]{Eubank:04}%
  \BibitemOpen
  \bibfield  {author} {\bibinfo {author} {\bibfnamefont {S.}~\bibnamefont
  {Eubank}}, \bibinfo {author} {\bibfnamefont {H.}~\bibnamefont {Guclu}},
  \bibinfo {author} {\bibfnamefont {V.~S.~A.}\ \bibnamefont {Kumar}}, \bibinfo
  {author} {\bibfnamefont {M.}~\bibnamefont {Marathe}}, \bibinfo {author}
  {\bibfnamefont {A.}~\bibnamefont {Srinivasan}}, \bibinfo {author}
  {\bibfnamefont {Z.}~\bibnamefont {Toroczkai}}, \ and\ \bibinfo {author}
  {\bibfnamefont {N.}~\bibnamefont {Wang}},\ }\href@noop {} {\bibfield
  {journal} {\bibinfo  {journal} {Nature}\ }\textbf {\bibinfo {volume}
  {429(6988)}},\ \bibinfo {pages} {180} (\bibinfo {year} {2004})}\BibitemShut
  {NoStop}%
\bibitem [{\citenamefont {Parikh}\ \emph {et~al.}(2013)\citenamefont {Parikh},
  \citenamefont {Youssef}, \citenamefont {Swarup},\ and\ \citenamefont
  {Eubank}}]{SocialNet}%
  \BibitemOpen
  \bibfield  {author} {\bibinfo {author} {\bibfnamefont {N.}~\bibnamefont
  {Parikh}}, \bibinfo {author} {\bibfnamefont {M.}~\bibnamefont {Youssef}},
  \bibinfo {author} {\bibfnamefont {S.}~\bibnamefont {Swarup}}, \ and\ \bibinfo
  {author} {\bibfnamefont {S.}~\bibnamefont {Eubank}},\ }\href@noop {}
  {\bibfield  {journal} {\bibinfo  {journal} {Scientific Reports}\ }\textbf
  {\bibinfo {volume} {3}} (\bibinfo {year} {2013})}\BibitemShut {NoStop}%
\bibitem [{\citenamefont {Valle}\ \emph {et~al.}(2007)\citenamefont {Valle},
  \citenamefont {Hyman}, \citenamefont {Hethcote},\ and\ \citenamefont
  {Eubank}}]{MixPattern}%
  \BibitemOpen
  \bibfield  {author} {\bibinfo {author} {\bibfnamefont {S.~Y.~D.}\
  \bibnamefont {Valle}}, \bibinfo {author} {\bibfnamefont {J.~M.}\ \bibnamefont
  {Hyman}}, \bibinfo {author} {\bibfnamefont {H.~W.}\ \bibnamefont {Hethcote}},
  \ and\ \bibinfo {author} {\bibfnamefont {S.~G.}\ \bibnamefont {Eubank}},\
  }\href@noop {} {\bibfield  {journal} {\bibinfo  {journal} {Social Network}\
  }\textbf {\bibinfo {volume} {29}},\ \bibinfo {pages} {539,554} (\bibinfo
  {year} {2007})}\BibitemShut {NoStop}%
\bibitem [{\citenamefont {Gao}\ \emph {et~al.}(2012)\citenamefont {Gao},
  \citenamefont {Buldyrev}, \citenamefont {Stanley},\ and\ \citenamefont
  {Havlin}}]{InterNet}%
  \BibitemOpen
  \bibfield  {author} {\bibinfo {author} {\bibfnamefont {J.}~\bibnamefont
  {Gao}}, \bibinfo {author} {\bibfnamefont {S.~V.}\ \bibnamefont {Buldyrev}},
  \bibinfo {author} {\bibfnamefont {H.~E.}\ \bibnamefont {Stanley}}, \ and\
  \bibinfo {author} {\bibfnamefont {S.}~\bibnamefont {Havlin}},\ }\href@noop {}
  {\bibfield  {journal} {\bibinfo  {journal} {Nat. Phys.}\ }\textbf {\bibinfo
  {volume} {8}},\ \bibinfo {pages} {40,48} (\bibinfo {year}
  {2012})}\BibitemShut {NoStop}%
\bibitem [{\citenamefont {Buldyrev}\ \emph {et~al.}(2009)\citenamefont
  {Buldyrev}, \citenamefont {Parshani}, \citenamefont {Paul}, \citenamefont
  {Stanley},\ and\ \citenamefont {Havlin}}]{Robustness}%
  \BibitemOpen
  \bibfield  {author} {\bibinfo {author} {\bibfnamefont {S.~V.}\ \bibnamefont
  {Buldyrev}}, \bibinfo {author} {\bibfnamefont {R.}~\bibnamefont {Parshani}},
  \bibinfo {author} {\bibfnamefont {G.}~\bibnamefont {Paul}}, \bibinfo {author}
  {\bibfnamefont {H.~E.}\ \bibnamefont {Stanley}}, \ and\ \bibinfo {author}
  {\bibfnamefont {S.}~\bibnamefont {Havlin}},\ }\href@noop {} {\bibfield
  {journal} {\bibinfo  {journal} {Nature}\ }\textbf {\bibinfo {volume} {464}},\
  \bibinfo {pages} {1025,1028} (\bibinfo {year} {2009})}\BibitemShut {NoStop}%
\bibitem [{\citenamefont {Trajanovski}\ \emph {et~al.}(2013)\citenamefont
  {Trajanovski}, \citenamefont {Martín-Hernández}, \citenamefont
  {Winterbach},\ and\ \citenamefont {{Van Mieghem}}}]{robustness1}%
  \BibitemOpen
  \bibfield  {author} {\bibinfo {author} {\bibfnamefont {S.}~\bibnamefont
  {Trajanovski}}, \bibinfo {author} {\bibfnamefont {J.}~\bibnamefont
  {Martín-Hernández}}, \bibinfo {author} {\bibfnamefont {W.}~\bibnamefont
  {Winterbach}}, \ and\ \bibinfo {author} {\bibfnamefont {P.}~\bibnamefont
  {{Van Mieghem}}},\ }\href@noop {} {\bibfield  {journal} {\bibinfo  {journal}
  {Journal of Complex Networks}\ }\textbf {\bibinfo {volume} {1}},\ \bibinfo
  {pages} {44} (\bibinfo {year} {2013})}\BibitemShut {NoStop}%
\bibitem [{\citenamefont {{Van Mieghem}}(2012)}]{robustness2}%
  \BibitemOpen
  \bibfield  {author} {\bibinfo {author} {\bibfnamefont {P.}~\bibnamefont {{Van
  Mieghem}}},\ }\href@noop {} {\bibfield  {journal} {\bibinfo  {journal}
  {Computer Communications}\ }\textbf {\bibinfo {volume} {35}},\ \bibinfo
  {pages} {1494} (\bibinfo {year} {2012})}\BibitemShut {NoStop}%
\bibitem [{\citenamefont {Youssef}\ \emph {et~al.}(2011)\citenamefont
  {Youssef}, \citenamefont {Kooij},\ and\ \citenamefont
  {Scoglio}}]{Youssef:11b}%
  \BibitemOpen
  \bibfield  {author} {\bibinfo {author} {\bibfnamefont {M.}~\bibnamefont
  {Youssef}}, \bibinfo {author} {\bibfnamefont {R.}~\bibnamefont {Kooij}}, \
  and\ \bibinfo {author} {\bibfnamefont {C.}~\bibnamefont {Scoglio}},\
  }\href@noop {} {\bibfield  {journal} {\bibinfo  {journal} {Journal of
  Computational Science}\ }\textbf {\bibinfo {volume} {2}},\ \bibinfo {pages}
  {286} (\bibinfo {year} {2011})}\BibitemShut {NoStop}%
\bibitem [{\citenamefont {Ellens}\ and\ \citenamefont
  {Kooij}(2013)}]{robustness4}%
  \BibitemOpen
  \bibfield  {author} {\bibinfo {author} {\bibfnamefont {W.}~\bibnamefont
  {Ellens}}\ and\ \bibinfo {author} {\bibfnamefont {R.}~\bibnamefont {Kooij}},\
  }\href@noop {} {\bibfield  {journal} {\bibinfo  {journal} {arXiv:1311.5064
  [cs.DM]}\ } (\bibinfo {year} {2013})}\BibitemShut {NoStop}%
\end{thebibliography}%

\end{document}